\documentclass[aps,prd,preprint,nofootinbib,11pt]{revtex4}
\usepackage{amsfonts}
\usepackage{mathrsfs}
\usepackage{graphicx}
\usepackage{amsmath}
\usepackage{amssymb}
\usepackage{subfigure}
\usepackage{epsfig}
\usepackage{graphicx}
\usepackage{ulem}
\usepackage{color}
\newcommand{\bqa}{\begin{eqnarray}}
\newcommand{\eqa}{\end{eqnarray}}
\newcommand{\beq}{\begin{equation}}
\newcommand{\eeq}{\end{equation}}
\allowdisplaybreaks[1]
\graphicspath{{fig/}{dia/}} \DeclareGraphicsExtensions{.eps}

\begin{document}
\title{\Large Light baryonium states with exotic quantum numbers\\[7mm]}

\author{Bing-Dong Wan$^{1,2}$\footnote{wanbd@lnnu.edu.cn}, Jun-Hao Zhang$^{1,2}$, Yan Zhang$^{1,2}$ and Ming-Yang Yuan$^{1,2}$ \vspace{+3pt}}

\affiliation{$^1$Department of Physics, Liaoning Normal University, Dalian 116029, China\\
$^2$ Center for Theoretical and Experimental High Energy Physics, Liaoning Normal University, Dalian 116029, China}

\author{~\\~\\}

\begin{abstract}
\vspace{0.3cm}
The existence of baryonium-bound or resonant states composed of a baryon and an antibaryon has long been postulated as a natural extension of conventional hadron spectroscopy. In the present work, we conduct a systematic investigation of the mass spectrum and internal configurations of light baryonium candidates exhibiting exotic quantum numbers that are inaccessible within the framework of the traditional quark model. Employing the method of QCD sum rules, we analyze nucleon–antinucleon and light hyperon–anti-hyperon systems with quantum numbers $J^{PC}=0^{--}$ and $0^{+-}$, which are quantum number combinations prohibited for conventional mesonic states. Our analysis reveals the potential existence of two $0^{--}$ $\Lambda$-$\bar{\Lambda}$ baryonium states with masses $(2.90\pm0.09)$ GeV and $(3.36\pm0.09)$ GeV, respectively, as well as two $0^{+-}$ $\Lambda$-$\bar{\Lambda}$ states with masses $(2.91\pm0.07)$ GeV and $(3.29\pm0.07)$ GeV, respectively. In addition, corresponding nucleon–antinucleon partner states are identified at $(2.69\pm0.07)$ GeV, $(3.07\pm0.08)$ GeV, $(2.86\pm0.07)$ GeV, and $(3.22\pm0.07)$ GeV, respectively. Furthermore, analogous $\Xi$-$\bar{\Xi}$ configurations are predicted with masses of $(3.10\pm0.09)$ GeV, $(3.54\pm0.07)$ GeV, $(3.08\pm0.08)$ GeV, and $(3.45\pm0.08)$ GeV, respectively. The possible decay modes of the light exotic baryonium states are analyzed, which are hopefully measurable in BESIII, BELLEII, and LHCb experiments.
\end{abstract}
 \maketitle
\newpage

\section{Introduction}
 Phenomenological studies of novel hadronic states offer fresh insights into hadron structure and the nature of confinement in QCD. The continuous interplay between experimental discoveries and theoretical advancements continues to deepen our understanding of hadron spectroscopy, shedding light on fundamental aspects of the strong interaction. In recent years, a series of so-called charmonium- and bottomonium-like XYZ states have been observed in experiments \cite{Choi:2003ue, Aubert:2005rm, Belle:2011aa, Ablikim:2013mio, Liu:2013dau}, providing new insights into the emergent structures of QCD.
 
In light of the experimental confirmation of tetraquark and pentaquark states, it is a natural and timely extension to hypothesize the existence of hexaquark states, and to initiate focused searches for such configurations. Among them, the deuteron, a bound state of a proton and neutron formed in the early Universe and its stability is responsible for the production of other elements, stands as a prototypical and experimentally well-established dibaryon molecular state with quantum numbers $J^P=1^+$ and a binding energy $E_B = 2.225 \text{MeV}$ \cite{Weinberg:1962hj}. Remarkably, the strong nuclear force that ensures the stability of the deuteron also suggests the theoretical possibility of a broader family of stable deuteronlike dibaryon states. However, despite longstanding theoretical interest and numerous proposals~\cite{Jaffe:1976yi,Mulders:1980vx,Balachandran:1983dj,Sakai:1999qm,Ikeda:2007nz,Bashkanov:2013cla,Shanahan:2011su,Clement:2016vnl}, no unambiguous experimental evidence for such states has yet been reported~\cite{BaBar:2018hpv}.

 In this context, baryonium states, which are composed of a baryon–antibaryon pair, represent a distinct class of hexaquark configurations. The interaction dynamics between a baryon and an antibaryon exhibit notable similarities to those between two baryons. However, baryon–antibaryon pairs are more readily produced in high-energy collider environments compared to dibaryon states. Furthermore, the presence of annihilation interactions in the baryon–antibaryon system often leads to more tightly bound configurations. Consequently, the investigation of baryonium states may provide critical insights into the mechanisms underlying the apparent non-observation of stable dibaryon states.
 
  Actually, the investigation into the baryon-antibaryon system traces back to the 1940s when Fermi and Yang proposed that $\pi$-mesons might be composite particles formed by a nucleon-antinucleon pair~\cite{Fermi:1949voc}, and their scenario was later on replaced by the quark
model. At the turn of the new millennium, heavy baryonium emerged as a novel hadronic state, was proposed as a potential explanation for the unusual properties of $Y(4260)$ \cite{Qiao:2005av, Qiao:2007ce} and other charmonium-like states observed in experiments. In the following years, research on baryonium expanded, with investigations conducted from various perspectives \cite{Chen:2011cta, Chen:2013sba, Wan:2019ake, Chen:2016ymy, Liu:2007tj, Wang:2021qmn, Wan:2022uie, Liu:2021gva, Wan:2021vny, Wan:2023epq,Zhang:2025qmg}.

In the light hadron spectrum, the small spacing between various states generally makes it difficult to distinguish novel hadronic states from conventional hadrons, except when the former exhibit unique quantum numbers. However, with a large sample of $J/\psi$ events, the BESIII Collaboration is carefully investigating the physics of the energy region around 2.0 GeV \cite{BESIII:2010gmv, BES:2003aic, BES:2005ega, BESIII:2010vwa, BESIII:2019wkp, BESIII:2016qzq, BESIII:2020vtu, BESIII:2017kqw, BESIII:2017hyw, BESIII:2019cuv}, reigniting interest in light novel hadronic states.
In our previous work, the mass of the light baryinium $\Lambda$-$\bar{\Lambda}$ state with $J^{PC}=1^{--}$ was calculated to be $(2.34\pm0.12)$ GeV~\cite{Wan:2021vny}. Later, by analyzing the process $e^+e^-\to \Lambda\bar{\Lambda}\eta$, BESIII Collaboration observed a $1^{--}$ structure in the $\Lambda\bar{\Lambda}$ invariant mass at $(2356\pm24)$ MeV with a decay width of $(304\pm82)$ MeV ~\cite{BESIII:2022tvj}. However, this structure is so broad that it is difficult to identify it as the $\Lambda$-$\bar{\Lambda}$ baryonium state, casting a shadow over the search for light baryonium. Fortunately, a set of light baryonium states possess exotic quantum numbers, such as $0^{--}$, $0^{+-}$, and so on. Since these quantum numbers are inaccessible to quark-antiquark bound states, they allow light baryonium states to be distinguished from mixed $q\bar{q}$ states.

In this work, the exotic light baryonium states with quantum numbers $J^{PC}=0^{--}$ and $0^{+-}$ are systematically investigated within the framework of QCD sum rules (QCDSR). QCDSR provide a robust nonperturbative approach that connects the underlying dynamics of QCD with hadronic phenomenology. Originally formulated by Shifman, Vainshtein, and Zakharov~\cite{Shifman}, this method relies on the analysis of two-point correlation functions constructed from carefully selected interpolating currents, which possess the same quantum numbers and partonic content (quarks and/or gluons) as the hadronic states under consideration. By employing the Operator Product Expansion (OPE), the short-distance (perturbative) and long-distance (nonperturbative) contributions to the correlation functions are systematically separated, with the latter encoded in terms of QCD vacuum condensates such as quark and gluon condensates. On the phenomenological side, the correlation function is expressed in terms of hadronic parameters, where the interpolating current is interpreted as an operator responsible for the creation and annihilation of the corresponding hadronic state. By equating the OPE representation with its phenomenological representation of the correlation functions, one derives the QCD sum rules, from which hadronic parameters—most notably the mass—can be extracted. The method has been extensively and effectively employed in the investigation of both conventional and exotic hadrons~\cite{Albuquerque:2013ija,Wang:2013vex,Govaerts:1984hc,Reinders:1984sr,P.Col,Narison:1989aq,Tang:2021zti,Qiao:2014vva,Qiao:2015iea,Tang:2019nwv,Wan:2020oxt,Wan:2022xkx,Zhang:2022obn,Wan:2024dmi,Tang:2024zvf,Li:2024ctd,Zhao:2023imq,Yin:2021cbb,Yang:2020wkh,Wan:2024cpc,Wan:2024pet,Wan:2024ykm,Zhang:2024jvv,Tang:2024kmh,Tang:2016pcf,Tang:2015twt,Qiao:2013dda,Qiao:2013raa,Wan:2020fsk,Wan:2025xhf,Chen:2014vha,Azizi:2019xla,Wang:2017sto,Wan:2025bdr,Wan:2025zau,Wan:2025ikc,Wan:2025sae,Tang:2025ept,Ben:2025wqn,Zhang:2025vqg,Zhang:2024ulk,Zhang:2024asb,Zhang:2024ick,Zhang:2023nxl}.

The rest of the paper is arranged as follows. After the introduction, a brief interpretation of QCD sum rules and some primary formulas in our calculation are presented in Sec. \ref{Formalism}. We using QCDSR to evaluate the masses of light baryonium states with $J^{PC}=0^{--}$ and $0^{+-}$ in Sec. \ref{Numerical}. The possible decay modes of the light exotic baryonium states are analyzed in Sec. \ref{decay}. The last part is left for a brief summary.

\section{A brief introduction to QCD sum rules}\label{Formalism}

The theoretical foundation of QCD sum rules is established through the analysis of the two-point correlation function, defined as
\begin{eqnarray}
\Pi(q^2) &=& i \int d^4 x e^{i q \cdot x} \langle 0 | T [ j(x), j^\dagger(0) ] | 0 \rangle\;,
\end{eqnarray}
where $j(x)$ denotes the interpolating current carrying the same quantum numbers and quark content as the hadronic state under consideration. The essential idea is to extract hadronic properties by comparing two distinct representations of this correlator: the theoretical (OPE) side and the phenomenological side.

On the OPE side, the correlation function is computed within the framework of QCD perturbation theory, supplemented by non-perturbative effects parameterized through vacuum condensates. In this approach, $\Pi(q^2)$ can be expressed via a dispersion relation as
\begin{eqnarray}
\Pi(q^2) &=& \int_{s_{\text{min}}}^{\infty} ds \frac{\rho^{\text{OPE}}(s)}{s - q^2} \; , \label{OPE-hadron}
\end{eqnarray}
where $s_{min}$ represents the kinematic threshold, typically given by the squared sum of the current-quark masses associated with the hadron \cite{Albuquerque:2013ija}, and $\rho^{OPE}(s) = \text{Im} [\Pi^{OPE}(s)] / \pi$  is the spectral density of the OPE side and contains the contributions of the condensates up to dimension 12 which can be expressed as:
\begin{eqnarray}
\rho^{OPE}(s) & = & \rho^{pert}(s) + \rho^{\langle \bar{q} q
\rangle}(s) +\rho^{\langle G^2 \rangle}(s) + \rho^{\langle \bar{q} G q \rangle}(s)
+ \rho^{\langle \bar{q} q \rangle^2}(s) + \rho^{\langle G^3 \rangle}(s) \nonumber\\
&+& \rho^{\langle \bar{q} q \rangle\langle \bar{q} G q \rangle}(s) +  \rho^{\langle \bar{q} q
\rangle^2\langle G^2 \rangle}(s) + \rho^{\langle \bar{q} G q \rangle^2}(s) +\rho^{\langle \bar{q} q \rangle^2\langle \bar{q} G q \rangle}(s)+ \rho^{\langle \bar{q} q
\rangle^4}(s)   \nonumber\\
&=& \sum_n c_n s^n  . \label{rho-OPE}
\end{eqnarray}
Here, $c_n$ is the coefficient of the $s^n$ term with corresponding vacuum condensates.

 To calculate the spectral density of the OPE side, Eq. (\ref{rho-OPE}), the full propagators $S^q_{i j}(x)$ of a light quark ($q=u$, $d$ or $s$) are used:
\begin{eqnarray}
S^q_{j k}(x) \! \! & = & \! \! \frac{i \delta_{j k} x\!\!\!\slash}{2 \pi^2
x^4} - \frac{\delta_{jk} m_q}{4 \pi^2 x^2} - \frac{i t^a_{j k} G^a_{\alpha\beta}}{32 \pi^2 x^2}(\sigma^{\alpha \beta} x\!\!\!\slash
+ x\!\!\!\slash \sigma^{\alpha \beta}) - \frac{\delta_{jk}}{12} \langle\bar{q} q \rangle + \frac{i\delta_{j k}
x\!\!\!\slash}{48} m_q \langle \bar{q}q \rangle - \frac{\delta_{j k} x^2}{192} \langle g_s \bar{q} \sigma \cdot G q \rangle \nonumber \\ &+& \frac{i \delta_{jk} x^2 x\!\!\!\slash}{1152} m_q \langle g_s \bar{q} \sigma \cdot G q \rangle - \frac{t^a_{j k} \sigma_{\alpha \beta}}{192}
\langle g_s \bar{q} \sigma \cdot G q \rangle
+ \frac{i t^a_{jk}}{768} (\sigma_{\alpha \beta} x \!\!\!\slash + x\!\!\!\slash \sigma_{\alpha \beta}) m_q \langle
g_s \bar{q} \sigma \cdot G q \rangle \;,
\end{eqnarray}
where, the vacuum condensates are clearly displayed. For more explanation on above propagator, readers may refer to Refs.~\cite{Wang:2013vex, Albuquerque:2013ija}.

On the phenomenological side, the same correlator is modeled using a spectral function that isolates the pole contribution of the ground state from those of higher excitations and the continuum. This representation takes the form
\begin{eqnarray}
\Pi^{\text{phen}}(q^2) &=& \frac{\lambda^2}{M^2 - q^2} + \frac{1}{\pi} \int_{s_0}^{\infty} ds \frac{\rho(s)}{s - q^2} \; , \label{hadron}
\end{eqnarray}
where $M$ is the mass of the hadronic state, $\lambda$ is the pole residue (coupling constant), and $s_0$ denotes the continuum threshold above which the spectral density $\rho(s)$ accounts for the contributions from excited and continuum states.

Invoking the quark–hadron duality hypothesis, the OPE and phenomenological representations are equated:
\begin{eqnarray}
\int_{s_{\text{min}}}^{\infty} ds \frac{\rho^{\text{OPE}}(s)}{s - q^2} &=& \frac{\lambda^2}{M^2 - q^2} + \frac{1}{\pi} \int_{s_0}^{\infty} ds \frac{\rho(s)}{s - q^2} \; , \label{mainSR}
\end{eqnarray}
thereby establishing the QCD sum rule relation. Applying a Borel transformation to Eq. (\ref{mainSR}) suppresses the continuum contributions and enhances the ground-state signal, ultimately allowing the extraction of the hadron mass through
\begin{eqnarray}
M(s_0, M_B^2) &=& \sqrt{ - \frac{L_1(s_0, M_B^2)}{L_0(s_0, M_B^2)} } \; , \label{mass-Eq}
\end{eqnarray}
where $M_B^2$ denotes the Borel parameter. The auxiliary functions $L_0$ and $L_1$ are defined as
\begin{eqnarray}
L_0(s_0, M_B^2) &=& \int_{s_{\text{min}}}^{s_0} ds \, \rho^{\text{OPE}}(s) \, e^{-s/M_B^2} \; , \label{L0}
\end{eqnarray}
\begin{eqnarray}
L_1(s_0, M_B^2) &=& \frac{\partial}{\partial (1/M_B^2)} L_0(s_0, M_B^2) \; .
\end{eqnarray}

\section{Exotic baryonium states via QCDSR}\label{Numerical}

In this work, the interpolating currents are constructed starting from baryonic degrees of freedom. We first build color-singlet baryon currents according to the quantum numbers and quark contents of the corresponding baryons. These baryonic currents are then treated as effective spinor fields, and combined with their antibaryon counterparts to form baryon--antibaryon type interpolating currents.

At the quark level, such constructions correspond to local six-quark operators. From a physical point of view, they naturally describe baryon--antibaryon molecular-like configurations. Therefore, the interpolating currents employed in this work provide a suitable framework to investigate baryonium states with exotic quantum numbers.

The interpolating currents for $0^{--}$ $\Lambda$-$\bar{\Lambda}$ baryonium state can be constructed as follow:
\begin{eqnarray}\label{current_0--}
j_{0^{--}}^A(x)&=&\frac{\epsilon_{a b c}\epsilon_{d e f}}{\sqrt{2}} {\Big\{}[\bar{s}_d(x) s_c(x)][q^T_a(x) C q^\prime_b(x)][\bar{q}_e (x)\gamma_5 C \bar{q}^{\prime T}_f(x)] \nonumber\\
&-&[\bar{s}_d(x) s_c(x)][q^T_a(x) C \gamma_5 q^\prime_b(x)][\bar{q}_e (x) C \bar{q}^{\prime T}_f(x)] {\Big\}}\;,\label{Ja0--}\\
j_{0^{--}}^B(x)&=&\frac{\epsilon_{a b c}\epsilon_{d e f}}{\sqrt{2}}{\Big\{}[\bar{s}_d(x) s_c(x)][q^T_a(x) C \gamma_\mu q^\prime_b(x)][\bar{q}_e(x) \gamma_\mu\gamma_5 C \bar{q}^{\prime T}_f(x)] \nonumber\\
&-&[\bar{s}_d(x) s_c(x)][q^T_a(x) C \gamma_\mu\gamma_5 q^\prime_b(x)][\bar{q}_e (x) \gamma_\mu C \bar{q}^{\prime T}_f(x)] {\Big\}}\;,\label{Jb0--}\\
j_{0^{--}}^C(x)&=& \frac{\epsilon_{a b c}\epsilon_{d e f}}{\sqrt{2}} {\Big\{}[\bar{s}_d(x) \gamma_\mu s_c(x)][q^T_a(x) C \gamma_\mu q^\prime_b(x)][\bar{q}_e (x) \gamma_5 C \bar{q}^{\prime T}_f (x)]\nonumber\\
&+&[\bar{s}_d(x) \gamma_\mu s_c(x)][q^T_a(x) C \gamma_5 q^\prime_b(x)][\bar{q}_e (x) \gamma_\mu C \bar{q}^{\prime T}_f (x)] {\Big\}}\;,\label{Jc0--}\\
j_{0^{--}}^D(x)&=&\frac{\epsilon_{a b c}\epsilon_{d e f}}{\sqrt{2}} {\Big\{}[\bar{s}_d(x) \gamma_\mu s_c(x)][q^T_a(x) C \gamma_\mu\gamma_5 q^\prime_b(x)][\bar{q}_e (x) C \bar{q}^{\prime T}_f (x)]\nonumber\\
&+&[\bar{s}_d(x) \gamma_\mu s_c(x)][q^T_a(x) C q^\prime_b(x)][\bar{q}_e (x) \gamma_\mu\gamma_5 C \bar{q}^{\prime T}_f (x)] {\Big\}}\;, \label{Jd0--}\\
j_{0^{--}}^E(x)&=& \frac{\epsilon_{a b c}\epsilon_{d e f}}{\sqrt{2}} {\Big\{}[\bar{s}_d(x) \gamma_\mu\gamma_5 s_c(x)][q^T_a(x) C \gamma_\mu\gamma_5 q^\prime_b(x)][\bar{q}_e (x) \gamma_5 C \bar{q}^{\prime T}_f (x)]\nonumber\\
&-&[\bar{s}_d(x) \gamma_\mu\gamma_5 s_c(x)][q^T_a(x) C \gamma_5 q^\prime_b(x)][\bar{q}_e (x) \gamma_\mu\gamma_5 C \bar{q}^{\prime T}_f (x)] {\Big\}}\;,\label{Je0--}\\
j_{0^{--}}^F(x)&=&\frac{\epsilon_{a b c}\epsilon_{d e f}}{\sqrt{2}}{\Big\{} [\bar{s}_d(x)  \gamma_\mu\gamma_5 s_c(x)][q^T_a(x) C \gamma_\mu q^\prime_b(x)][\bar{q}_e (x) C \bar{q}^{\prime T}_f (x)]\nonumber\\
&-& [\bar{s}_d(x)  \gamma_\mu\gamma_5 s_c(x)][q^T_a(x) C q^\prime_b(x)][\bar{q}_e (x) \gamma_\mu C \bar{q}^{\prime T}_f (x)] {\Big\}}\;. \label{Jf0--}
\end{eqnarray}
For the $0^{+-}$ $\Lambda$-$\bar{\Lambda}$ states, the interpolating currents are found to be in forms:
\begin{eqnarray}\label{current_0+-}
j_{0^{+-}}^A(x)&=&\frac{\epsilon_{a b c}\epsilon_{d e f}}{\sqrt{2}} {\Big\{}[\bar{s}_d(x) \gamma_5 s_c(x)][q^T_a(x) C q^\prime_b(x)][\bar{q}_e (x)\gamma_5 C \bar{q}^{\prime T}_f(x)] \nonumber\\
&-&[\bar{s}_d(x) \gamma_5 s_c(x)][q^T_a(x) C \gamma_5 q^\prime_b(x)][\bar{q}_e (x) C \bar{q}^{\prime T}_f(x)] {\Big\}}\;,\label{Ja0+-}\\
j_{0^{+-}}^B(x)&=&\frac{\epsilon_{a b c}\epsilon_{d e f}}{\sqrt{2}}{\Big\{}[\bar{s}_d(x)\gamma_5 s_c(x)][q^T_a(x) C \gamma_\mu q^\prime_b(x)][\bar{q}_e(x) \gamma_\mu\gamma_5 C \bar{q}^{\prime T}_f(x)] \nonumber\\
&-&[\bar{s}_d(x) \gamma_5 s_c(x)][q^T_a(x) C \gamma_\mu\gamma_5 q^\prime_b(x)][\bar{q}_e (x) \gamma_\mu C \bar{q}^{\prime T}_f(x)] {\Big\}}\;,\label{Jb0+-}\\
j_{0^{+-}}^C(x)&=& \frac{\epsilon_{a b c}\epsilon_{d e f}}{\sqrt{2}} {\Big\{}[\bar{s}_d(x) \gamma_\mu s_c(x)][q^T_a(x) C \gamma_\mu q^\prime_b(x)][\bar{q}_e (x) C \bar{q}^{\prime T}_f (x)]\nonumber\\
&+&[\bar{s}_d(x) \gamma_\mu s_c(x)][q^T_a(x) C q^\prime_b(x)][\bar{q}_e (x) \gamma_\mu C \bar{q}^{\prime T}_f (x)] {\Big\}}\;,\label{Jc0+-}\\
j_{0^{+-}}^D(x)&=&\frac{\epsilon_{a b c}\epsilon_{d e f}}{\sqrt{2}} {\Big\{}[\bar{s}_d(x) \gamma_\mu s_c(x)][q^T_a(x) C \gamma_\mu\gamma_5 q^\prime_b(x)][\bar{q}_e (x) \gamma_5 C \bar{q}^{\prime T}_f (x)]\nonumber\\
&+&[\bar{s}_d(x) \gamma_\mu s_c(x)][q^T_a(x) C \gamma_5 q^\prime_b(x)][\bar{q}_e (x) \gamma_\mu\gamma_5 C \bar{q}^{\prime T}_f (x)] {\Big\}}\;, \label{Jd0+-}\\
j_{0^{+-}}^E(x)&=& \frac{\epsilon_{a b c}\epsilon_{d e f}}{\sqrt{2}} {\Big\{}[\bar{s}_d(x) \gamma_\mu\gamma_5 s_c(x)][q^T_a(x) C \gamma_\mu q^\prime_b(x)][\bar{q}_e (x) \gamma_5 C \bar{q}^{\prime T}_f (x)]\nonumber\\
&-&[\bar{s}_d(x) \gamma_\mu\gamma_5 s_c(x)][q^T_a(x) C \gamma_5 q^\prime_b(x)][\bar{q}_e (x) \gamma_\mu C \bar{q}^{\prime T}_f (x)] {\Big\}}\;,\label{Je0+-}\\
j_{0^{+-}}^F(x)&=&\frac{\epsilon_{a b c}\epsilon_{d e f}}{\sqrt{2}}{\Big\{} [\bar{s}_d(x)  \gamma_\mu\gamma_5 s_c(x)][q^T_a(x) C \gamma_\mu\gamma_5 q^\prime_b(x)][\bar{q}_e (x) C \bar{q}^{\prime T}_f (x)]\nonumber\\
&-& [\bar{s}_d(x)  \gamma_\mu\gamma_5 s_c(x)][q^T_a(x) C q^\prime_b(x)][\bar{q}_e (x) \gamma_\mu\gamma_5 C \bar{q}^{\prime T}_f (x)] {\Big\}}\;. \label{Jf0+-}
\end{eqnarray}

Here, the subscripts $a\cdots f$ are color indices, $q$ and $q^\prime$ stand for light quark $u$ and $d$, respectively, and $C$ is the charge conjugation matrix.

We note that the notation ``$\Lambda$-$\bar{\Lambda}$ baryonium'' used in this work refers to the baryon--antibaryon nature of the system at the hadronic level. This naming is motivated by the construction of the interpolating currents from baryonic currents and their antibaryon counterparts.

At the same time, it should be emphasized that the interpolating currents are local operators at the quark level, corresponding to six-quark configurations. Therefore, the notation does not imply a simple product of a baryon and an antibaryon, nor does it necessarily correspond to a loosely bound molecular state in a spatial sense. Instead, it reflects the underlying flavor structure and baryon--antibaryon configuration of the system.

With the currents (\ref{Ja0--})-(\ref{Jf0+-}), the two-point correlation function can be readily established, and spectral density $\rho(s)$ can be calculated analytically.

In performing the numerical calculation, the broadly accepted inputs are taken \cite{Shifman,Albuquerque:2013ija,P.Col,Tang:2019nwv,ParticleDataGroup:2024cfk,Reinders:1984sr,Narison:1989aq}, i.e., $m_u=2.16^{+0.49}_{-0.26}\; \text{MeV}$, $m_d=4.67^{+0.48}_{-0.17}\; \text{MeV}$, $m_s=(95\pm5)\; \text{MeV}$, $\langle \bar{q} q \rangle = - (0.23 \pm 0.03)^3 \; \text{GeV}^3$, $\langle \bar{s} s \rangle=(0.8\pm0.1)\langle \bar{q} q \rangle$, $\langle \bar{q} g_s \sigma \cdot G q \rangle = m_0^2 \langle\bar{q} q \rangle$, $\langle \bar{s} g_s \sigma \cdot G s \rangle = m_0^2 \langle\bar{s} s \rangle$, $\langle g_s^2 G^2 \rangle = (0.88\pm0.25) \; \text{GeV}^4$, and $m_0^2 = (0.8 \pm 0.2) \; \text{GeV}^2$.

In this work, we retain finite values for the $u$ and $d$ quark masses, following the standard inputs from the Particle Data Group. Although the chiral limit ($m_u = m_d = 0$) is often adopted in QCD sum rule analyses, the inclusion of finite light-quark masses provides a more complete treatment. Numerically, we have checked that their effects are very small and do not affect the final results within the uncertainties.

Furthermore, the formulation of QCD sum rules introduces two additional parameters: the continuum threshold $s_0$ and the Borel parameter $M_B^2$. These parameters are determined according to the standard procedures outlined in Refs.~\cite{Shifman,Reinders:1984sr,P.Col}, which rely on two well-established criteria.
The first criterion concerns the convergence of the operator product expansion (OPE). Specifically, the relative magnitude of each higher-dimensional term is compared to the total OPE contribution, and an acceptable range of $M_B^2$ is selected to ensure that the truncated OPE remains convergent.
The second criterion is based on the pole contribution (PC). As noted in Refs.~\cite{Chen:2014vha,Azizi:2019xla,Wang:2017sto}, the presence of large powers of $s$ in the spectral density tends to suppress the PC value. For hexaquark states, the pole contribution is therefore required to exceed $15\%$ of the total, ensuring that the ground-state signal is sufficiently dominant. These two criteria can be expressed mathematically as follows:
\begin{eqnarray}
  R^{X,\;OPE}_{J^{PC}} = \left| \frac{L_{J^{PC},\;0}^{X,\;c_0}(s_0, M_B^2)}{L_{J^{PC},\;0}^X(s_0, M_B^2)}\right|\, ,
\end{eqnarray}
\begin{eqnarray}
  R^{X,\;PC}_{J^{PC}} = \frac{L_{J^{PC},\;0}^X(s_0, M_B^2)}{L_{J^{PC},\;0}^X(\infty, M_B^2)} \; , \label{RatioPC}
\end{eqnarray}
where the superscript $X$ runs from $A$ to $E$ denotes .

To determine an appropriate value for the continuum threshold $s_0$, we follow an analysis analogous to that in Refs.\cite{Qiao:2013dda,Tang:2016pcf,Qiao:2013raa}. The objective is to identify a value of $s_0$ that yields an optimal stability window for the mass curve of the baryonium state. Within this window, the extracted mass should exhibit minimal dependence on the Borel parameter $M_B^2$. In practical implementation, $s_0$ is varied by $0.1$ GeV to determine its lower and upper bounds, thereby providing an estimate of the uncertainty in $s_0$ \cite{Wan:2020oxt,Wan:2020fsk}.

\begin{figure}
\includegraphics[width=6.8cm]{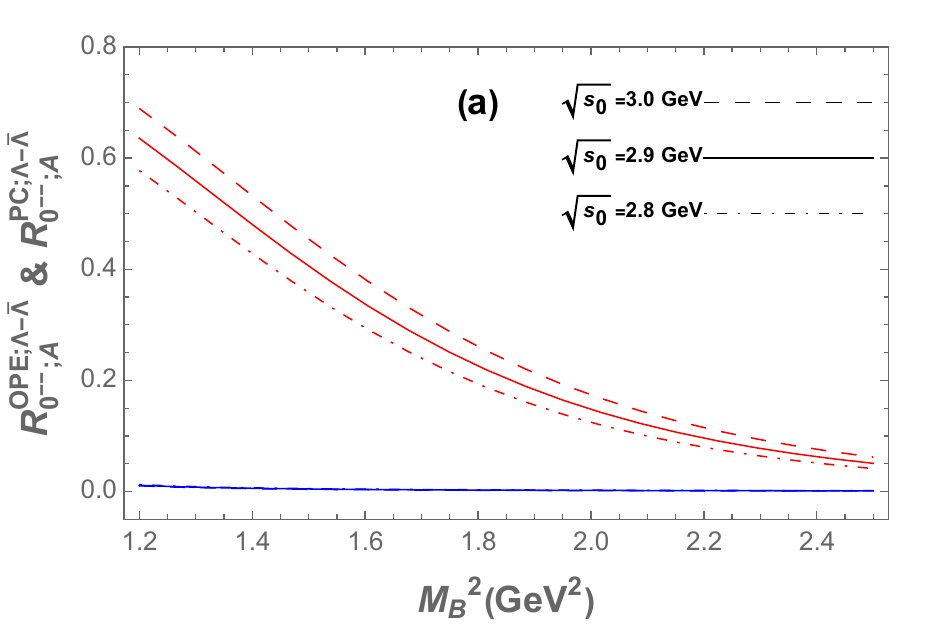}
\includegraphics[width=6.8cm]{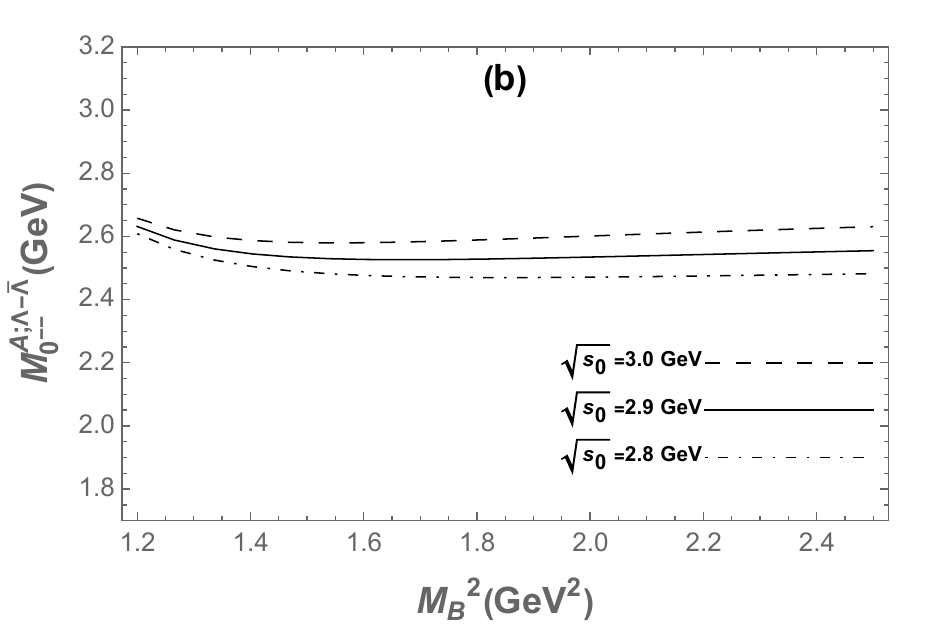}
\caption{ (a) The ratios of $R^{OPE\;,A}_{0^{--}}$ and $R^{PC\;,A}_{0^{--}}$ as functions of the Borel parameter $M_B^2$ for different values of $\sqrt{s_0}$, where blue lines represent $R^{OPE\;,A}_{0^{--}}$ and red lines denote $R^{PC\;,A}_{0^{--}}$. (b) The mass $M^{A}_{0^{--}}$ as a function of the Borel parameter $M_B^2$ for different values of $\sqrt{s_0}$.} \label{figA0--}
\end{figure}

With the above preparation the mass spectrum of exotic baryonium states can be numerically evaluated. For the $0^{--}$ $\Lambda$-$\bar{\Lambda}$ baryonium state in Eq.~(\ref{Ja0--}), the ratios $R^{OPE\;,A}_{0^{--}}$ and $R^{PC\;,A}_{0^{--}}$ are presented as functions of Borel parameter $M_B^2$ in Fig. \ref{figA0--}(a) with different values of $\sqrt{s_0}$, i.e., $3.1$, $3.2$, and $3.3$ GeV. The reliant relations of $M^{A}_{0^{--}}$ on parameter $M_B^2$ are displayed in Fig. \ref{figA0--}(b). The optimal Borel window is found in range $1.4 \le M_B^2 \le 2.1\; \text{GeV}^2$, and the mass $M^{A}_{0^{--}}$ can then be obtained:
\begin{eqnarray}
M^{A}_{0^{--}} &=& (2.90\pm 0.09)\; \text{GeV}.\label{m1}
\end{eqnarray}

\begin{figure}[h]
\includegraphics[width=6.8cm]{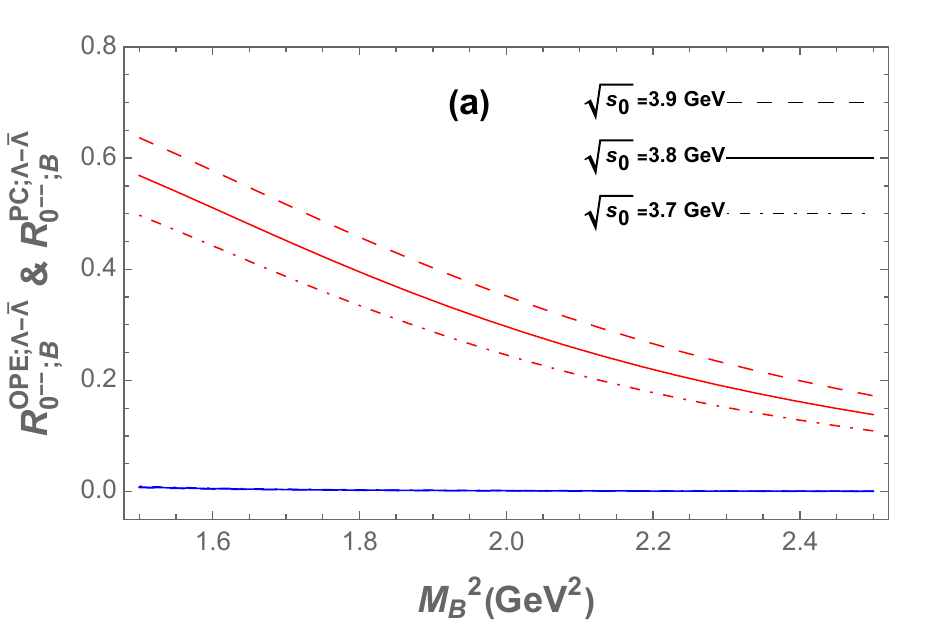}
\includegraphics[width=6.8cm]{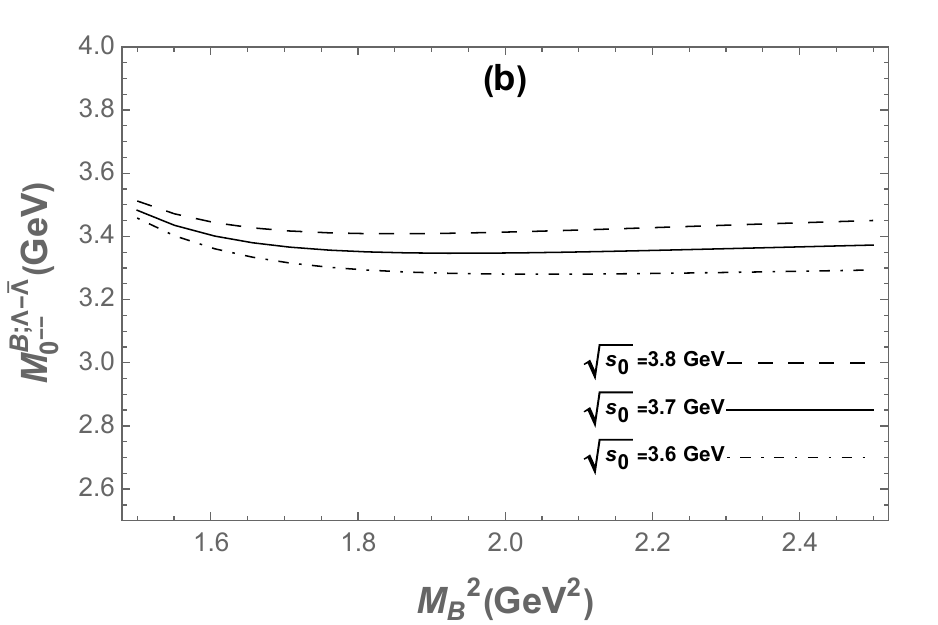}
\caption{(a) The ratios of $R^{OPE\;,B}_{0^{--}}$ and $R^{PC\;,B}_{0^{--}}$ as functions of the Borel parameter $M_B^2$ for different values of $\sqrt{s_0}$, where blue lines represent $R^{OPE\;,B}_{0^{--}}$ and red lines denote $R^{PC\;,B}_{0^{--}}$. (b) The mass $M^{B}_{0^{--}}$ as a function of the Borel parameter $M_B^2$ for different values of $\sqrt{s_0}$.} \label{figB0--}
\end{figure}

For the $0^{--}$ $\Lambda$-$\bar{\Lambda}$ baryonium state defined in Eq.~(\ref{Jb0--}), the ratios $R^{OPE\;,B}_{0^{--}}$ and $R^{PC\;,B}_{0^{--}}$ are shown in Fig. \ref{figB0--}(a) as functions of Borel parameter $M_B^2$  with different values of $\sqrt{s_0}$, namely $3.6$, $3.7$, and $3.8$ GeV. The dependence of the extracted mass $M^{B}_{0^{--}}$ on $M_B^2$ is presented in Fig.~\ref{figB0--}(b). From this analysis, the optimal Borel window is determined to be $1.6 \le M_B^2 \le 2.4\; \text{GeV}^2$, within which the mass $M^{B}_{0^{--}}$ is obtained as:
\begin{eqnarray}
M^{B}_{0^{--}} &=& (3.36\pm 0.08)\; \text{GeV}.\label{m2}
\end{eqnarray}

\begin{figure}[h]
\includegraphics[width=6.8cm]{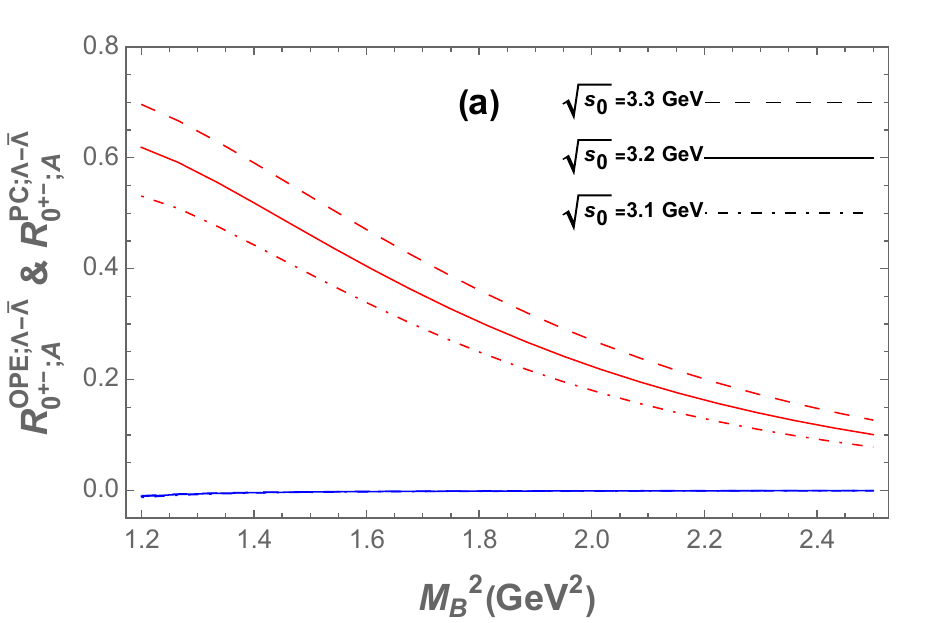}
\includegraphics[width=6.8cm]{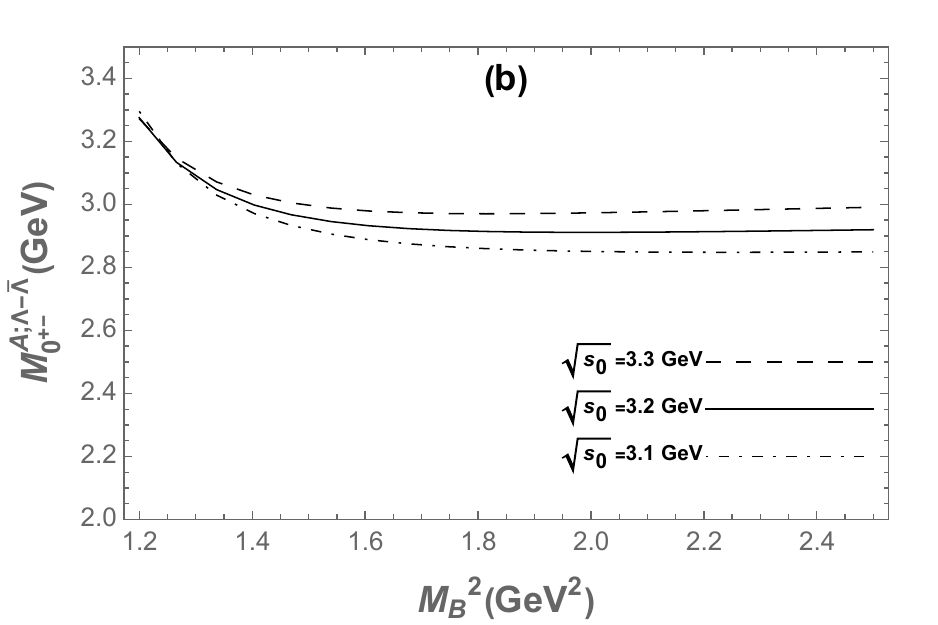}
\caption{(a) The ratios of $R^{OPE\;,A}_{0^{+-}}$ and $R^{PC\;,A}_{0^{+-}}$ as functions of the Borel parameter $M_B^2$ for different values of $\sqrt{s_0}$, where blue lines represent $R^{OPE\;,A}_{0^{+-}}$ and red lines denote $R^{PC\;,A}_{0^{+-}}$. (b) The mass $M^{A}_{0^{+-}}$ as a function of the Borel parameter $M_B^2$ for different values of $\sqrt{s_0}$.} \label{figA0+-}
\end{figure}

For the $0^{+-}$ $\Lambda$-$\bar{\Lambda}$ baryonium state defined in Eq.~(\ref{Ja0+-}), the ratios $R^{OPE\;,A}_{0^{+-}}$ and $R^{PC\;,A}_{0^{+-}}$ are displayed in Fig. \ref{figA0+-}(a) as functions of Borel parameter $M_B^2$ with different values of $\sqrt{s_0}$, namely $3.1$, $3.2$, and $3.3$ GeV. The dependence of the extracted mass $M^{A}_{0^{+-}}$ on $M_B^2$ is shown in Fig.~\ref{figA0+-}(b). Based on this analysis, the optimal Borel window is determined to be $1.6 \le M_B^2 \le 2.3\; \text{GeV}^2$, within which the mass $M^{A}_{0^{+-}}$ is extracted as:
\begin{eqnarray}
M^{A}_{0^{+-}} &=& (2.91\pm 0.07)\; \text{GeV}.\label{m3}
\end{eqnarray}

\begin{figure}[h]
\includegraphics[width=6.8cm]{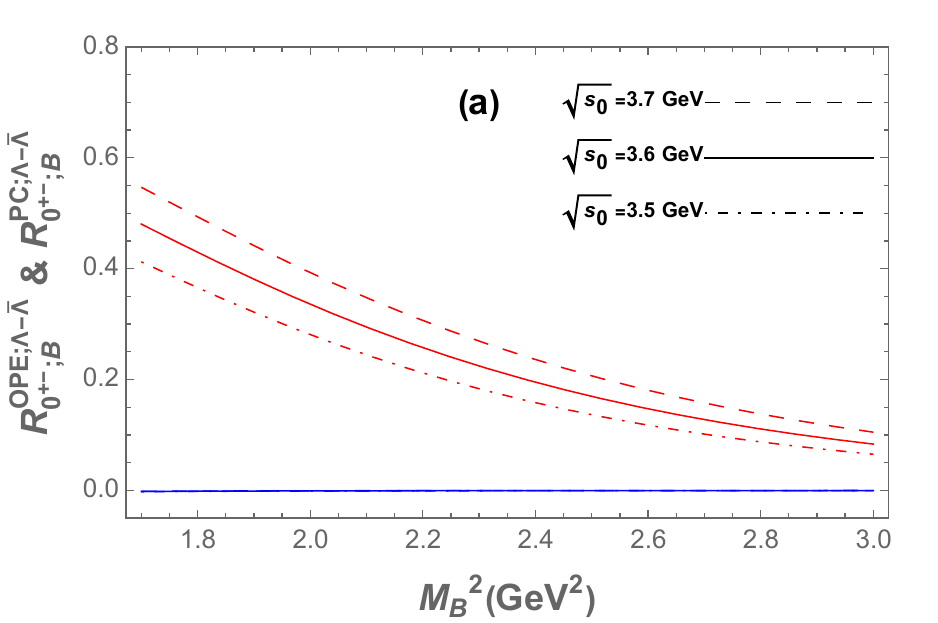}
\includegraphics[width=6.8cm]{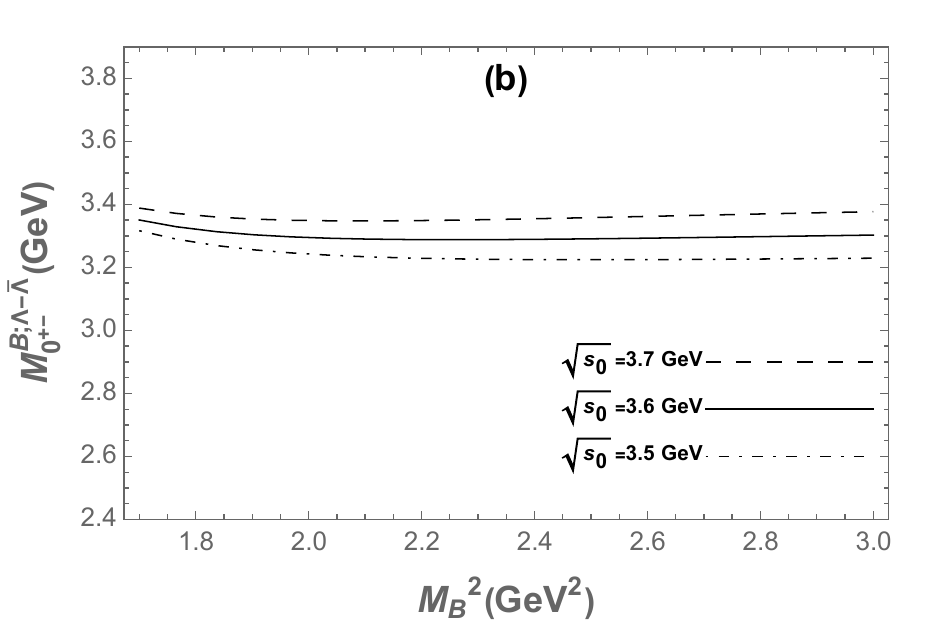}
\caption{(a) The ratios of $R^{OPE\;,B}_{0^{+-}}$ and $R^{PC\;,B}_{0^{+-}}$ as functions of the Borel parameter $M_B^2$ for different values of $\sqrt{s_0}$, where blue lines represent $R^{OPE\;,B}_{0^{+-}}$ and red lines denote $R^{PC\;,B}_{0^{+-}}$. (b) The mass $M^{B}_{0^{+-}}$ as a function of the Borel parameter $M_B^2$ for different values of $\sqrt{s_0}$.} \label{figB0+-}
\end{figure}

For the $0^{+-}$ $\Lambda$-$\bar{\Lambda}$ baryonium state defined in Eq.~(\ref{Jb0+-}), the ratios $R^{OPE\;,B}_{0^{+-}}$ and $R^{PC\;,B}_{0^{+-}}$ are illustrated in Fig. \ref{figB0+-}(a) as functions of Borel parameter $M_B^2$ for several representative values of $\sqrt{s_0}$, specifically $3.5$, $3.6$, and $3.7$ GeV. The dependence of the extracted mass $M^{B}_{0^{+-}}$ on $M_B^2$ is depicted in Fig.~\ref{figB0+-}(b). From this analysis, the optimal Borel window is determined to lie within the range $1.9 \le M_B^2 \le 2.6\; \text{GeV}^2$, within which the mass $M^{B}_{0^{+-}}$ is extracted as:
\begin{eqnarray}
M^{B}_{0^{+-}} &=& (3.29\pm 0.07)\; \text{GeV}.\label{m4}
\end{eqnarray}

The uncertainties in the results predominantly arise from the imprecisions in the quark masses, vacuum condensates, and the continuum threshold parameter $\sqrt{s_0}$. For the convenience of reference, a collection of Borel parameters, continuum thresholds, and predicted masses are tabulated in Table~\ref{mass}.

For the interpolating currents $j_{0^{--}}^C$, $j_{0^{--}}^D$, $j_{0^{--}}^E$, $j_{0^{--}}^F$, $j_{0^{+-}}^C$, $j_{0^{+-}}^D$, $j_{0^{+-}}^E$, and $j_{0^{+-}}^F$, no stable Borel window exhibiting a plateau is observed regardless of the chosen values of $M_B^2$ and $\sqrt{s_0}$. This indicates that the currents defined in Eqs.~(\ref{Jc0--}), (\ref{Jd0--}), (\ref{Je0--}), (\ref{Jf0--}), (\ref{Jc0+-}), (\ref{Jd0+-}), (\ref{Je0+-}), and (\ref{Jf0+-}) do not reliably couple to the corresponding baryonium states. 

By applying the same analysis while substituting the $s$-quark with either the $q$- or $q^\prime$-quark, one obtains the exotic $N$-$\bar{N}$ baryonium states. Conversely, replacing the $q$- or $q^\prime$-quark with $s$-quark, and simultaneously substituting the $s$-quark with the $q$- or $q^\prime$-quark, leads to the exotic $\Xi$-$\bar{\Xi}$ baryonium states. The corresponding OPE, pole contributions, and extracted masses as functions of the Borel parameter $M_B^2$ are presented in Appendix~\ref{pictures}. The resulting masses are summarized in Table~\ref{mass}.

\begin{table}
\begin{center}
\renewcommand\tabcolsep{10pt}
\caption{The continuum thresholds, Borel parameters, and predicted masses of exotic light baryonium.}\label{mass}
\begin{tabular}{cccccc}\hline\hline
States                                          &$J^{PC}$      &Current   & $\sqrt{s_0}\;(\text{GeV})$     &$M_B^2\;(\text{GeV}^2)$ &$M^X\;(\text{GeV})$       \\ \hline
$\Lambda-\bar{\Lambda}$           &$0^{--}$        &$A$        & $3.2\pm0.1$                             &$1.4-2.1$                      &$2.90\pm0.09$         \\
                                                    &                       &$B$        & $3.7\pm0.1$                             &$1.6-2.4$                      &$3.36\pm0.08$          \\\hline
$\Lambda-\bar{\Lambda}$          &$0^{+-}$       &$A$        & $3.2\pm0.1$                           &$1.6-2.3$                     &$2.91\pm0.07$           \\
                                                   &                      &$B$        & $3.6\pm0.1$                           &$1.9-2.6$                     &$3.29\pm0.07$           \\\hline
$N-\bar{N}$                                 &$0^{--}$        &$A$        & $3.0\pm0.1$                             &$1.5-2.0$                      &$2.69\pm0.07$         \\
                                                   &                    &$B$        & $3.4\pm0.1$                             &$1.6-2.2$                      &$3.07\pm0.08$          \\\hline
$N-\bar{N}$                                &$0^{+-}$       &$A$        & $3.1\pm0.1$                           &$1.6-2.3$                     &$2.86\pm0.07$           \\
                                                   &                      &$B$        & $3.5\pm0.1$                           &$1.9-2.6$                     &$3.22\pm0.07$           \\\hline   
$\Xi-\bar{\Xi}$                             &$0^{--}$        &$A$        & $3.5\pm0.1$                             &$1.8-2.5$                      &$3.10\pm0.09$         \\
                                                   &                    &$B$        & $3.8\pm0.1$                             &$1.9-2.5$                      &$3.54\pm0.07$          \\\hline       
$\Xi-\bar{\Xi}$                             &$0^{+-}$        &$A$        & $3.4\pm0.1$                             &$1.5-2.4$                      &$3.08\pm0.08$         \\
                                                   &                    &$B$        & $3.8\pm0.1$                             &$1.9-2.8$                      &$3.45\pm0.08$          \\\hline                                                                                                                                                                      
 \hline
\end{tabular}
\end{center}
\end{table}

\section{Decay analyses}\label{decay}

To unambiguously identify these exotic light baryonium states, the most direct approach involves reconstructing them from their decay products. However, a comprehensive understanding of their detailed properties requires further investigation. The representative decay channels of such exotic baryonium states are listed in Table \ref{decay}, and these processes are anticipated to be accessible in ongoing experiments at BESIII, Belle II, and the LHCb.

\begin{table}
\begin{center}\caption{Typical decay modes of the exotic light baryomium for each quantum number.}\label{decay}
\renewcommand\tabcolsep{8.0pt}
\begin{tabular}{cccccccc}
\hline
\hline
$J^{PC}$                                   &$0^{--}$                                                                                              &$0^{+-}$                                           \\\hline
$N$-$\bar{N}$                           & $N\Bar{N}(1535)$\;\;$N(1535)\Bar{N}$                                            &$N\Bar{N}$\;\;$N(1440)\Bar{N}(1440)$\\              
                                                  & $N\Bar{N}(1650)$\;\;$N(1650)\Bar{N}$                                             &$N(1520)\Bar{N}(1520)$\;\;$N(1535)\Bar{N}(1535)$ \\
		                                 & $N(1440)\Bar{N}(1535)$\;\;$N(1535)\Bar{N}(1440)$                         &$N(1650)\Bar{N}(1650)$\;\;$N\Bar{N}(1440)$  \\
		                                 &                                                                                                            &$N(1440)\Bar{N}$\;\;$N(1535)\Bar{N}(1650)$\\
		                                 &                                                                                                            &$N\Bar{N}(1710)$\;\;$N(1710)\Bar{N}$\\
		                                 &                                                                                                            & $N(1710)\Bar{N}(1440)$\;\;$N(1440)\Bar{N}(1710)$\\\hline
 $\Lambda$-$\bar{\Lambda}$   & $\Lambda\Bar{\Lambda}(1405)$\;\;$\Lambda(1405)\Bar{\Lambda}$  &$\Lambda\Bar{\Lambda}$\;\;$\Lambda(1405)\Bar{\Lambda}(1405)$  \\
      &$\Lambda\Bar{\Lambda}(1670)$\;\;$\Lambda(1670)\Bar{\Lambda}$    &$\Lambda(1520)\Bar{\Lambda}(1520)$\;\;$\Lambda(1600)\Bar{\Lambda}(1600)$       \\
       &$\Lambda(1600)\Bar{\Lambda}(1405)$\;\;$\Lambda(1405)\Bar{\Lambda}(1600)$ &$\Lambda(1600)\Bar{\Lambda}(1405)$\;\;$\Lambda(1405)\Bar{\Lambda}(1600)$ \\
       &$\Lambda(1600)\Bar{\Lambda}(1670)$\;\;$\Lambda(1670)\Bar{\Lambda}(1600)$ &$\Lambda\Bar{\Lambda}(1600)$\;\;$\Lambda(1600)\Bar{\Lambda}$ \\
       &$\Lambda(1520)\Bar{\Lambda}(1890)$\;\;$\Lambda(1890)\Bar{\Lambda}(1520)$&$\Lambda(1670)\Bar{\Lambda}(1405)$\;\;$\Lambda(1405)\Bar{\Lambda}(1670)$\\
       &                                                                                                                                &$\Lambda(1690)\Bar{\Lambda}(1520)$\;\;$\Lambda(1520)\Bar{\Lambda}(1690)$\\
       &                                                                                                                                &$\Lambda(1800)\Bar{\Lambda}(1405)$\;\;$\Lambda(1405)\Bar{\Lambda}(1800)$\\
       &                                                                                                                                &$\Lambda\Bar{\Lambda}(1810)$\;\;$\Lambda(1810)\Bar{\Lambda}$ \\\hline
$\Xi$-$\bar{\Xi}$    &$\Xi(1530)\bar{\Xi}(1820)$\;\;$\Xi(1820)\bar{\Xi}(1530)$               & $\Xi\bar{\Xi}$\;\;$\Xi(1530)\bar{\Xi}(1530)$  \\
                               &                                                                                                       &  $\Xi(1820)\bar{\Xi}(1820)$  \\
\hline
\hline
\end{tabular}
\end{center}
\end{table}

\section{Summary}

In summary, we have conducted a systematic investigation of light exotic baryonium states with quantum numbers $J^{PC}=0^{--}$ and $0^{+-}$ within the framework of QCD sum rules. The numerical results, presented in Table \ref{mass}, suggest the possible existence of two $0^{--}$ $\Lambda$-$\bar{\Lambda}$ baryonium states with masses of $(2.90\pm0.09)$ GeV and $(3.36\pm0.09)$ GeV, as well as two $0^{+-}$ $\Lambda$-$\bar{\Lambda}$ states with masses of $(2.91\pm0.07)$ GeV and $(3.29\pm0.07)$ GeV, respectively. Corresponding nucleon–antinucleon partner states are identified at $(2.69\pm0.07)$ GeV, $(3.07\pm0.08)$ GeV, $(2.86\pm0.07)$ GeV, and $(3.22\pm0.07)$ GeV, respectively. In addition, the analysis predicts analogous $\Xi$-$\bar{\Xi}$ configurations with masses of $(3.10\pm0.09)$ GeV, $(3.54\pm0.07)$ GeV, $(3.08\pm0.08)$ GeV, and $(3.45\pm0.08)$ GeV, respectively. Potential decay channels of these light exotic baryonium states have been examined and tabulated in Table~\ref{decay}, and such processes are expected to be experimentally accessible at BESIII, Belle II, and LHCb.

\vspace{.5cm} {\bf Acknowledgments} \vspace{.5cm}

This work was supported in part by the National Natural Science Foundation of China under Grants 12575106 and 12147214, and Specific Fund of Fundamental Scientific Research Operating Expenses for Undergraduate Universities in Liaoning Province under Grants No. LJ212410165019.


\begin{widetext}
\appendix

\section{OPE, pole contributions, and extracted masses as functions of the Borel parameter $M_B^2$ for $N$-$\bar{N}$ and $\Xi$-$\bar{\Xi}$ baryonium states}\label{pictures}
The corresponding OPE, pole contributions, and extracted masses as functions of the Borel parameter $M_B^2$ for $N$-$\bar{N}$ and $\Xi$-$\bar{\Xi}$ baryonium states are presented in this Appendix.

\begin{figure}[h]
\includegraphics[width=6.8cm]{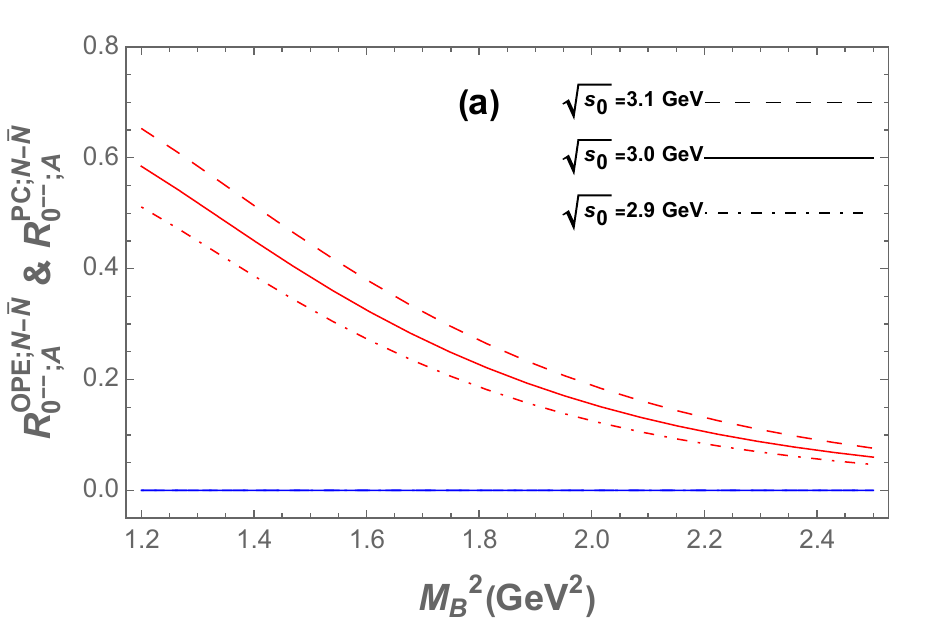}
\includegraphics[width=6.8cm]{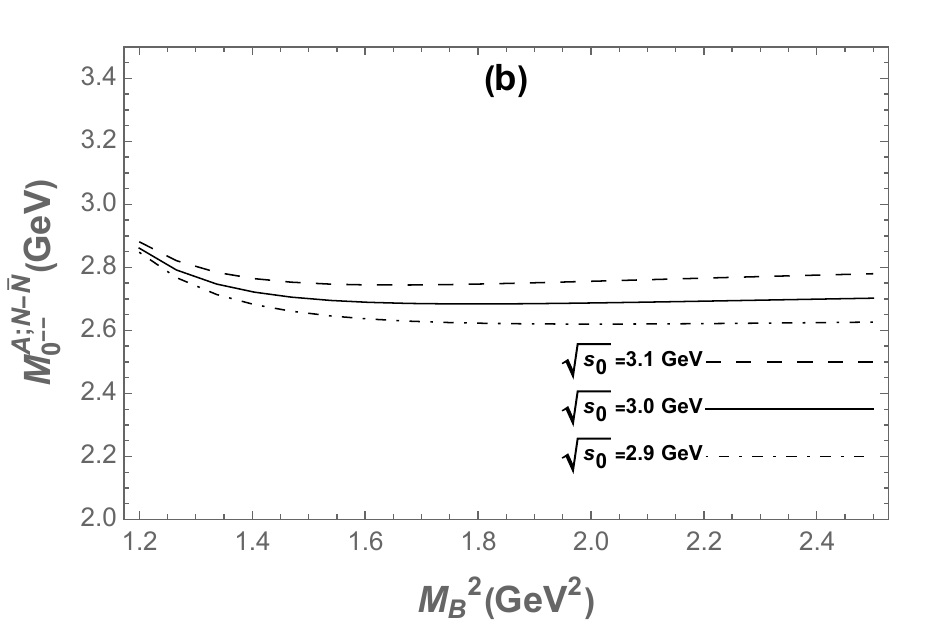}
\caption{(a) The ratios of $R^{OPE\;,A}_{0^{--},\;N-\bar{N}}$ and $R^{PC\;,A}_{0^{--},\;N-\bar{N}}$ as functions of the Borel parameter $M_B^2$ for different values of $\sqrt{s_0}$, where blue lines represent $R^{OPE\;,A}_{0^{--},\;N-\bar{N}}$ and red lines denote $R^{PC\;,A}_{0^{--},\;N-\bar{N}}$. (b) The mass $M^{A,\;N-\bar{N}}_{0^{--}}$ as a function of the Borel parameter $M_B^2$ for different values of $\sqrt{s_0}$.} \label{figA0--n}
\end{figure}

\begin{figure}[h]
\includegraphics[width=6.8cm]{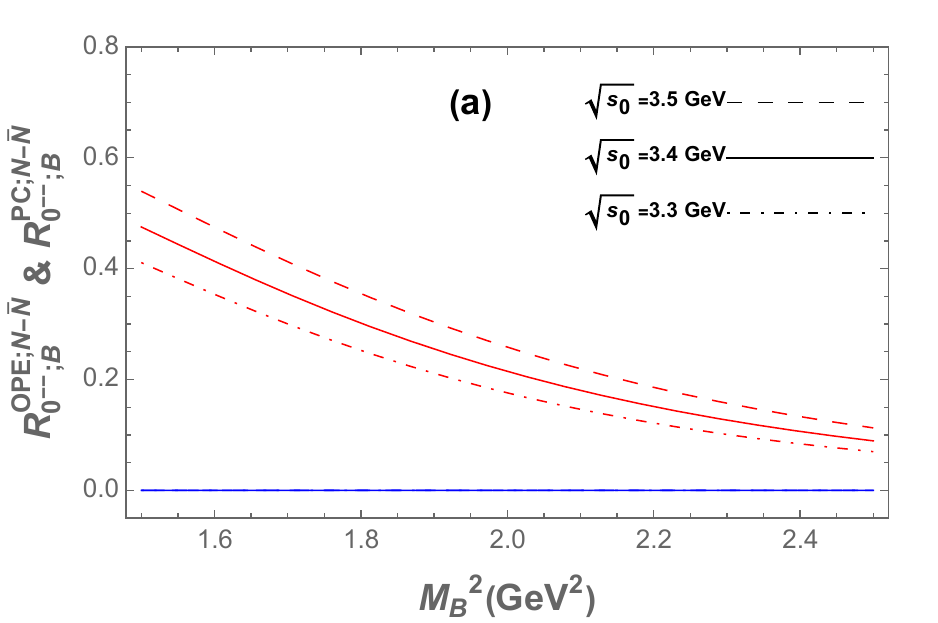}
\includegraphics[width=6.8cm]{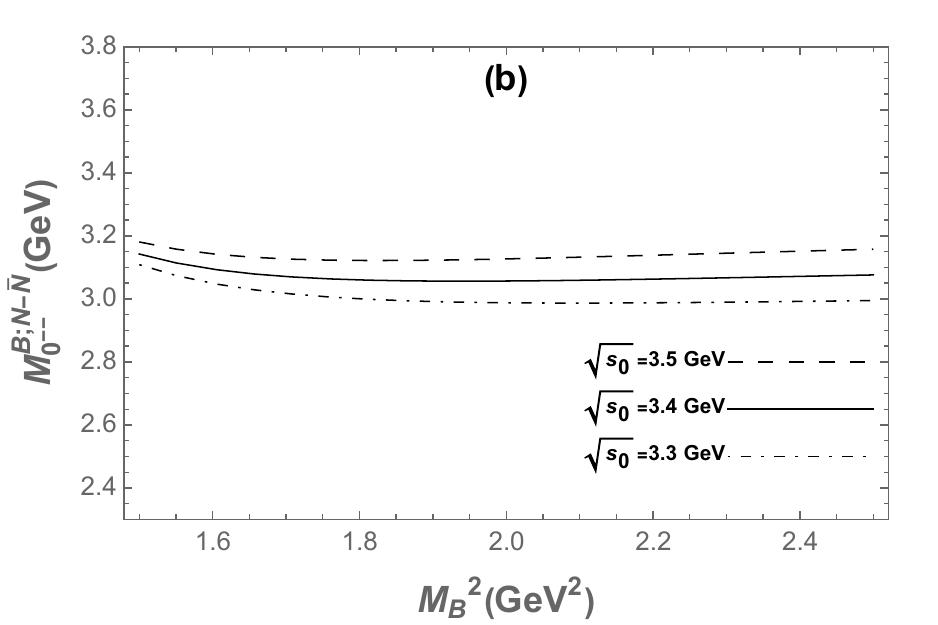}
\caption{(a) The ratios of $R^{OPE\;,B}_{0^{--},\;N-\bar{N}}$ and $R^{PC\;,B}_{0^{--},\;N-\bar{N}}$ as functions of the Borel parameter $M_B^2$ for different values of $\sqrt{s_0}$, where blue lines represent $R^{OPE\;,B}_{0^{--},\;N-\bar{N}}$ and red lines denote $R^{PC\;,B}_{0^{--},\;N-\bar{N}}$. (b) The mass $M^{B,\;N-\bar{N}}_{0^{--}}$ as a function of the Borel parameter $M_B^2$ for different values of $\sqrt{s_0}$.} \label{figB0--n}
\end{figure}

\begin{figure}[h]
\includegraphics[width=6.8cm]{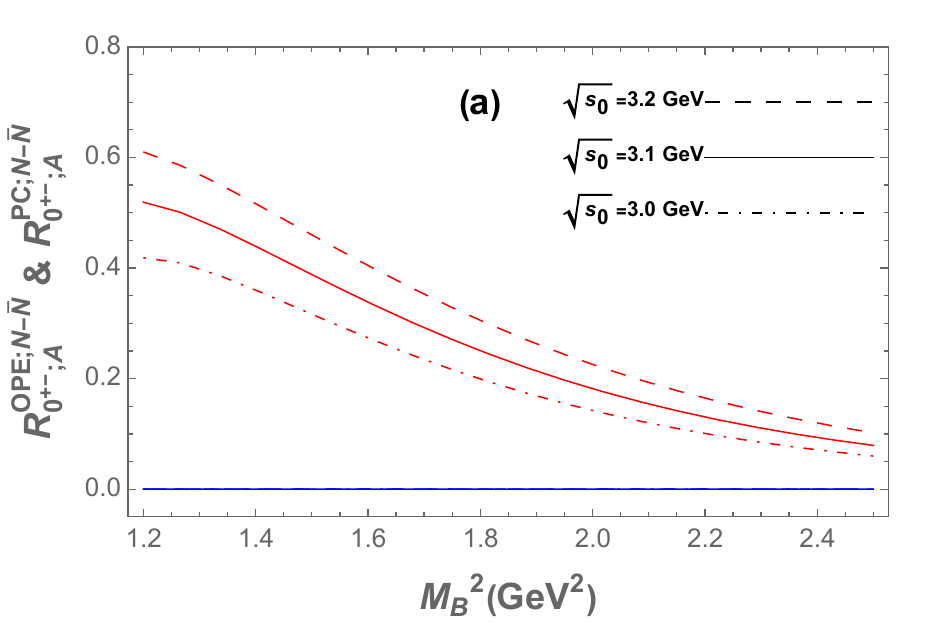}
\includegraphics[width=6.8cm]{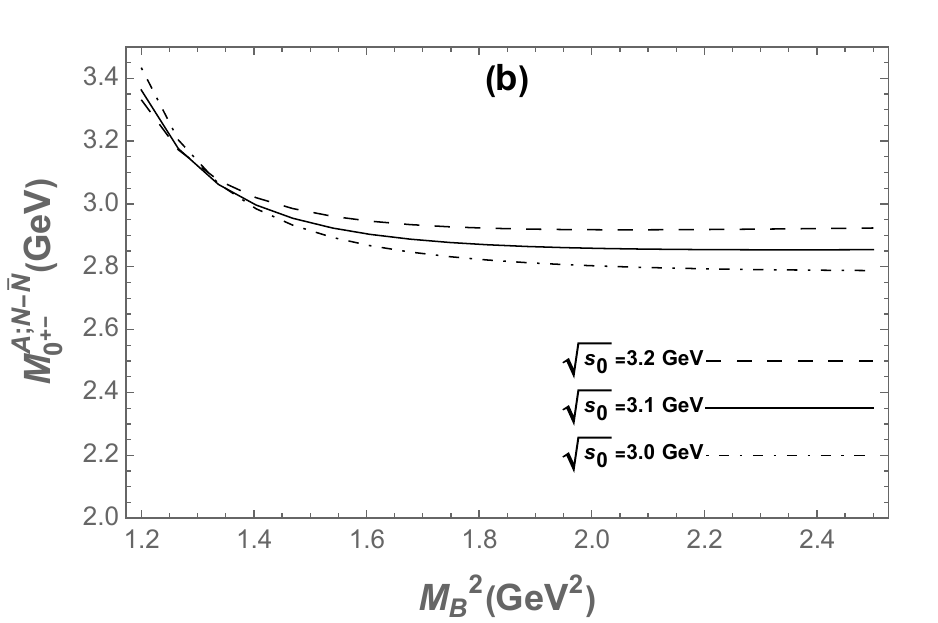}
\caption{(a) The ratios of $R^{OPE\;,A}_{0^{+-},\;N-\bar{N}}$ and $R^{PC\;,A}_{0^{+-},\;N-\bar{N}}$ as functions of the Borel parameter $M_B^2$ for different values of $\sqrt{s_0}$, where blue lines represent $R^{OPE\;,A}_{0^{+-},\;N-\bar{N}}$ and red lines denote $R^{PC\;,A}_{0^{+-},\;N-\bar{N}}$. (b) The mass $M^{A,\;N-\bar{N}}_{0^{+-}}$ as a function of the Borel parameter $M_B^2$ for different values of $\sqrt{s_0}$.} \label{figA0+-n}
\end{figure}

\begin{figure}[h]
\includegraphics[width=6.8cm]{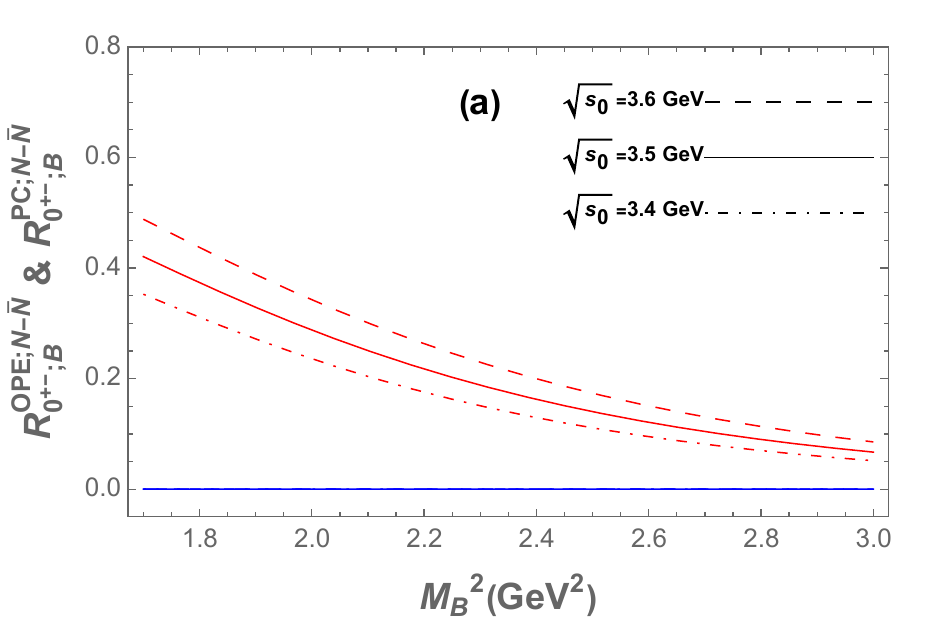}
\includegraphics[width=6.8cm]{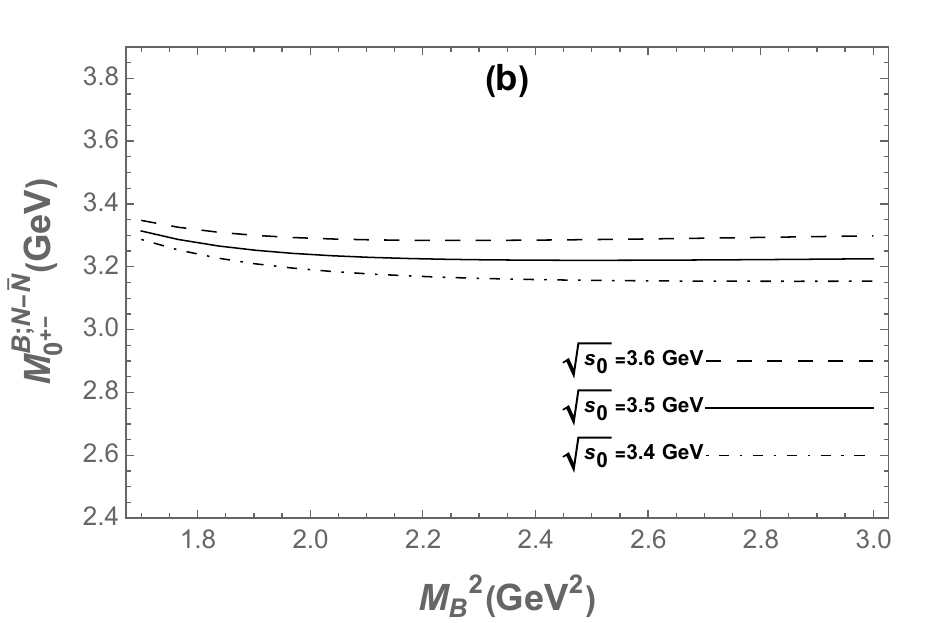}
\caption{(a) The ratios of $R^{OPE\;,B}_{0^{+-},\;N-\bar{N}}$ and $R^{PC\;,B}_{0^{+-},\;N-\bar{N}}$ as functions of the Borel parameter $M_B^2$ for different values of $\sqrt{s_0}$, where blue lines represent $R^{OPE\;,B}_{0^{+-},\;N-\bar{N}}$ and red lines denote $R^{PC\;,B}_{0^{+-},\;N-\bar{N}}$. (b) The mass $M^{B,\;N-\bar{N}}_{0^{+-}}$ as a function of the Borel parameter $M_B^2$ for different values of $\sqrt{s_0}$.} \label{figB0+-n}
\end{figure}

\begin{figure}[h]
\includegraphics[width=6.8cm]{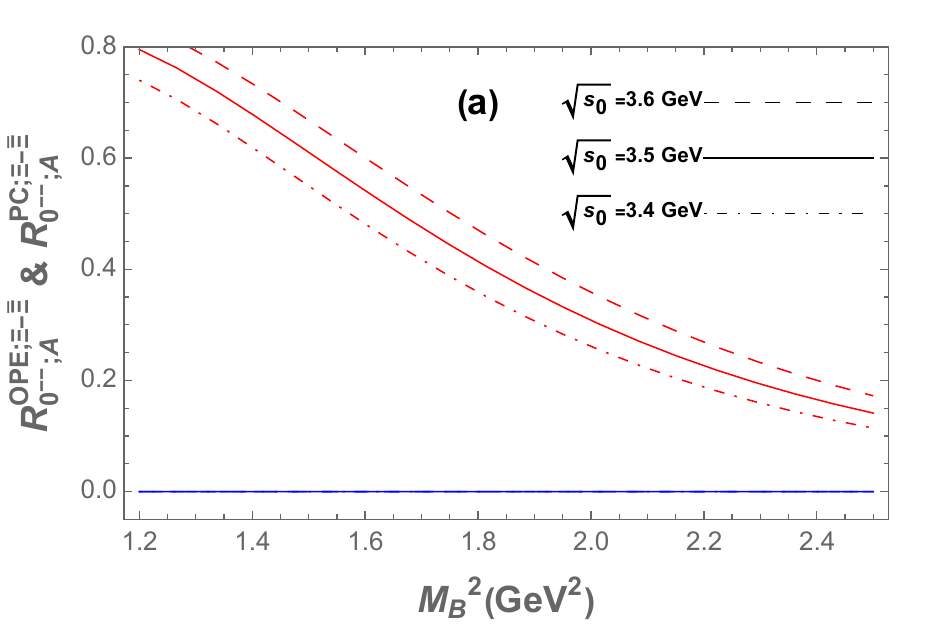}
\includegraphics[width=6.8cm]{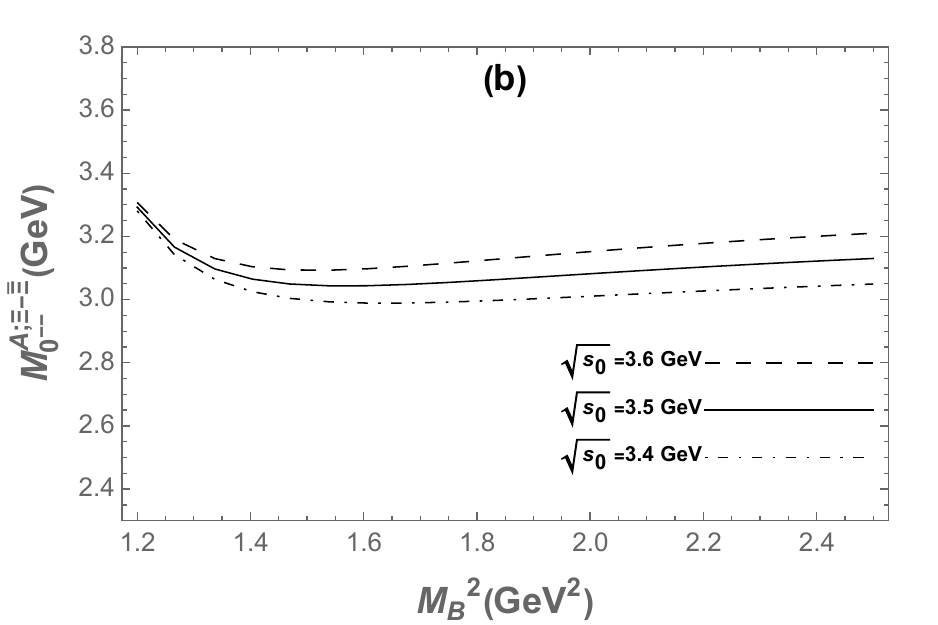}
\caption{(a) The ratios of $R^{OPE\;,A}_{0^{--},\;\Xi-\bar{\Xi}}$ and $R^{PC\;,A}_{0^{--},\;\Xi-\bar{\Xi}}$ as functions of the Borel parameter $M_B^2$ for different values of $\sqrt{s_0}$, where blue lines represent $R^{OPE\;,A}_{0^{--},\;\Xi-\bar{\Xi}}$ and red lines denote $R^{PC\;,A}_{0^{--},\;\Xi-\bar{\Xi}}$. (b) The mass $M^{A,\;\Xi-\bar{\Xi}}_{0^{--}}$ as a function of the Borel parameter $M_B^2$ for different values of $\sqrt{s_0}$.} \label{figA0--x}
\end{figure}

\begin{figure}[h]
\includegraphics[width=6.8cm]{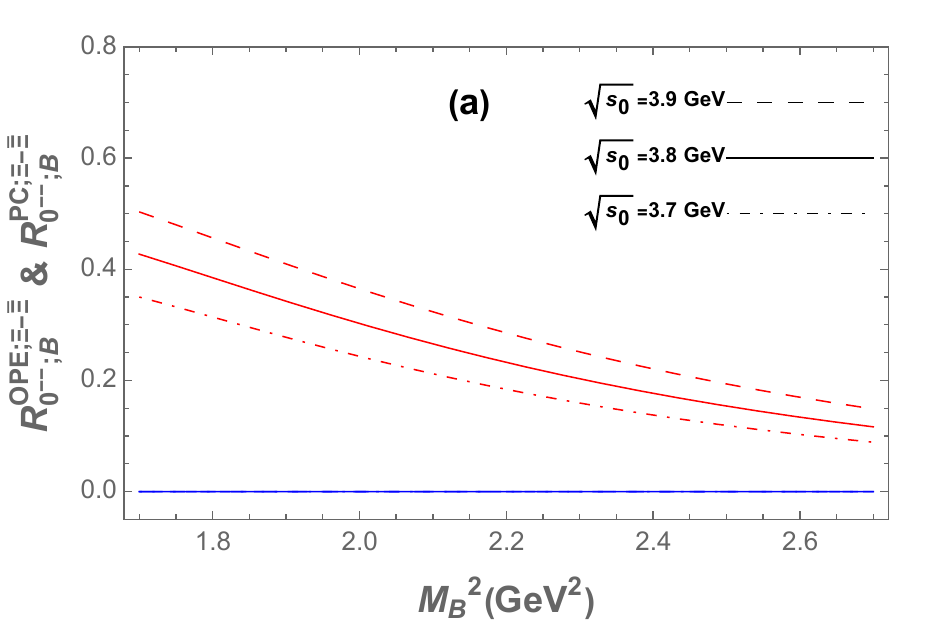}
\includegraphics[width=6.8cm]{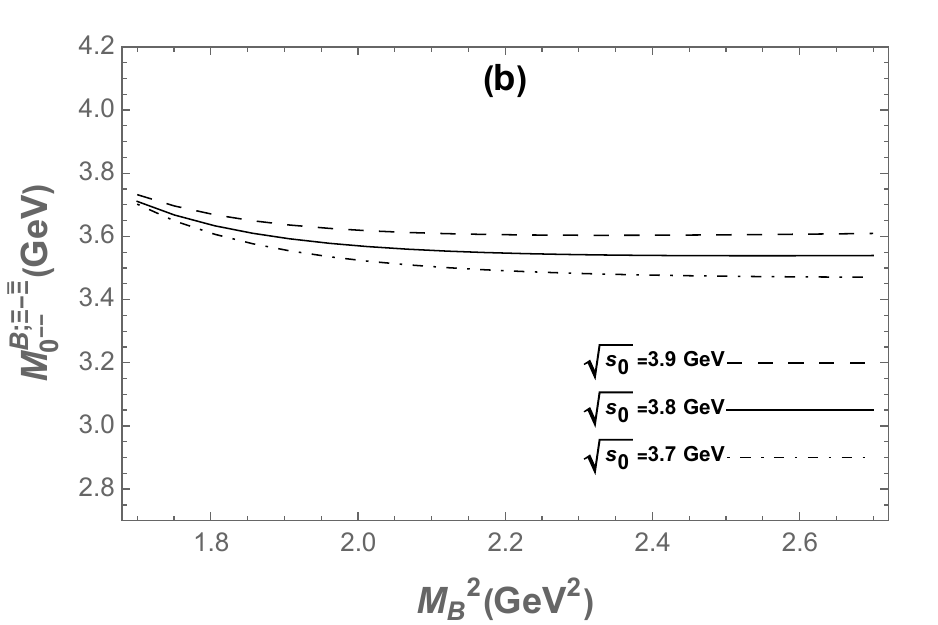}
\caption{(a) The ratios of $R^{OPE\;,B}_{0^{--},\;\Xi-\bar{\Xi}}$ and $R^{PC\;,B}_{0^{--},\;\Xi-\bar{\Xi}}$ as functions of the Borel parameter $M_B^2$ for different values of $\sqrt{s_0}$, where blue lines represent $R^{OPE\;,B}_{0^{--},\;\Xi-\bar{\Xi}}$ and red lines denote $R^{PC\;,B}_{0^{--},\;\Xi-\bar{\Xi}}$. (b) The mass $M^{B,\;\Xi-\bar{\Xi}}_{0^{--}}$ as a function of the Borel parameter $M_B^2$ for different values of $\sqrt{s_0}$.} \label{figB0--x}
\end{figure}

\begin{figure}[h]
\includegraphics[width=6.8cm]{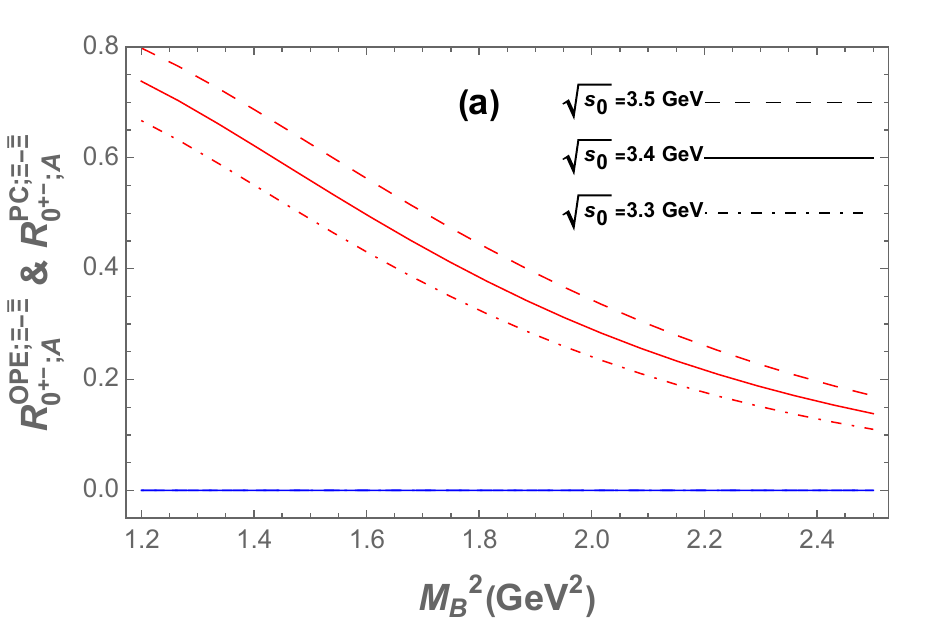}
\includegraphics[width=6.8cm]{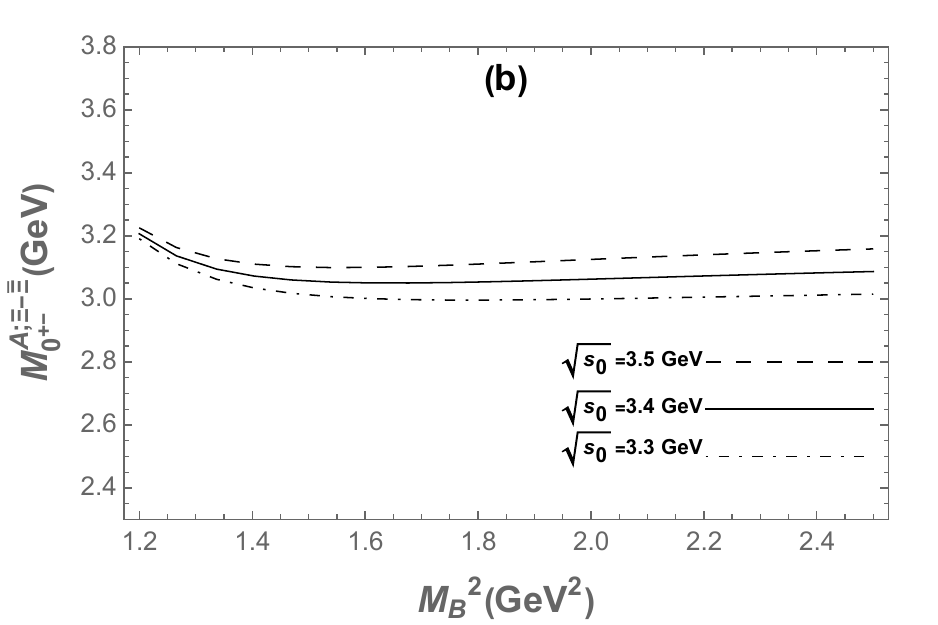}
\caption{(a) The ratios of $R^{OPE\;,A}_{0^{+-},\;\Xi-\bar{\Xi}}$ and $R^{PC\;,A}_{0^{+-},\;\Xi-\bar{\Xi}}$ as functions of the Borel parameter $M_B^2$ for different values of $\sqrt{s_0}$, where blue lines represent $R^{OPE\;,A}_{0^{+-},\;\Xi-\bar{\Xi}}$ and red lines denote $R^{PC\;,A}_{0^{+-},\;\Xi-\bar{\Xi}}$. (b) The mass $M^{A,\;\Xi-\bar{\Xi}}_{0^{+-}}$ as a function of the Borel parameter $M_B^2$ for different values of $\sqrt{s_0}$.} \label{figA0--x}
\end{figure}

\begin{figure}[h]
\includegraphics[width=6.8cm]{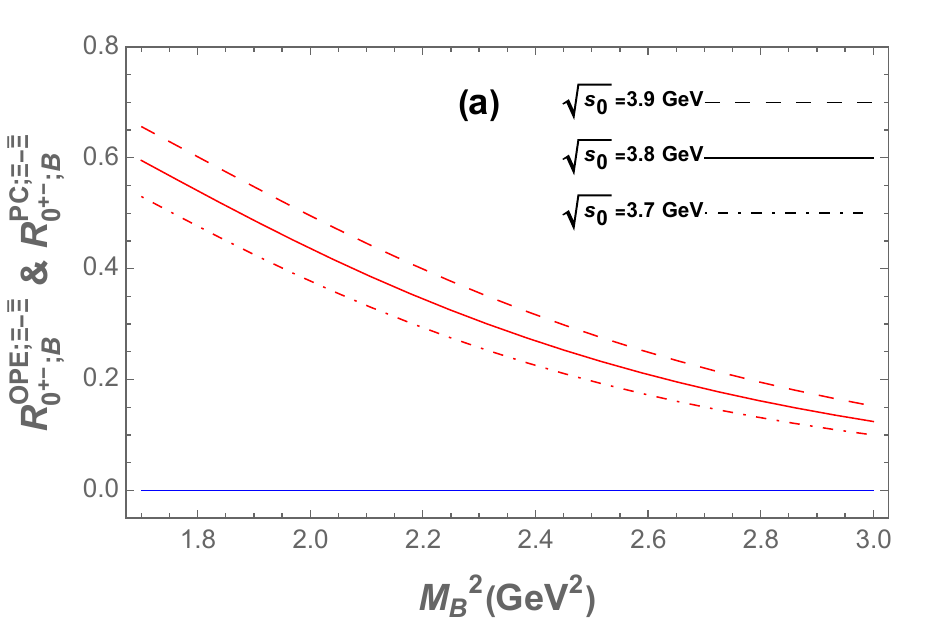}
\includegraphics[width=6.8cm]{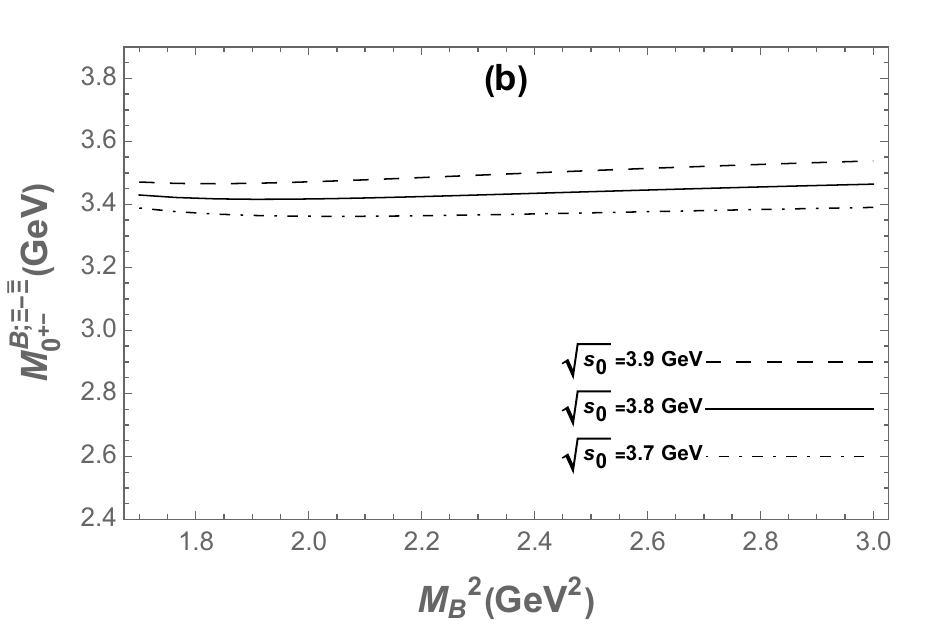}
\caption{(a) The ratios of $R^{OPE\;,B}_{0^{+-},\;\Xi-\bar{\Xi}}$ and $R^{PC\;,B}_{0^{+-},\;\Xi-\bar{\Xi}}$ as functions of the Borel parameter $M_B^2$ for different values of $\sqrt{s_0}$, where blue lines represent $R^{OPE\;,B}_{0^{+-},\;\Xi-\bar{\Xi}}$ and red lines denote $R^{PC\;,B}_{0^{+-},\;\Xi-\bar{\Xi}}$. (b) The mass $M^{B,\;\Xi-\bar{\Xi}}_{0^{+-}}$ as a function of the Borel parameter $M_B^2$ for different values of $\sqrt{s_0}$.} \label{figB0--x}
\end{figure}

\end{widetext}
\end{document}